\documentstyle[12pt,epsf]{article}
\begin{document}
\begin{titlepage}

\newcommand\letterhead {%
\hfill\parbox{8cm}{    \large \it UNIVERSITY of PENNSYLVANIA\\
\large Department of Physics\\
David Rittenhouse Laboratory\\
Philadelphia PA 19104--6396}\\[0.5cm]
{\bf PREPRINT UPR--0087NT}\\ July 1993}
\noindent\letterhead\par
\vspace{2 cm}
\noindent
\begin{center}
{\Large \bf Calculation of the properties of the rotational bands of
$^{155,157}\,$Gd}\\
\end{center}
\begin{center}
{\bf \large
Pavlos Protopapas, Abraham Klein, and Niels R. Walet}
\end{center}
\begin{center}
\vfill
{\em Submitted to Physical Review C}
\end{center}
\end{titlepage}

\title{Calculation of the properties of the rotational bands of
$^{155,157}\,$Gd}
\date{\today}
\author{Pavlos Protopapas, Abraham Klein, and
Niels R. Walet\thanks{address after Sept.~1, 1993:
Instit\"ut f\"ur theoretische Physik III, Universit\"at
Erlangen-N\"urnberg,
D-91058 Erlangen, Germany}}
\maketitle

\date{\today}
\begin{abstract}
We reexamine the long-standing problem of
the microscopic derivation of a particle-core coupling model.
We base our research on the Klein-Kerman approach, as amended
by D\"onau and Frauendorf. We describe the formalism to calculate
energy spectra and transition strengths in some detail.
We apply our formalism  to the rotational nuclei $^{155,157}$Gd,
where recent experimental data requires an explanation. We find no
clear evidence of a need for Coriolis attenuation.
\end{abstract}

\section{Introduction}

In this paper we use the Kerman-Klein method
\cite{kk:1,KleinReview,KleinWalet}
to derive a microscopic core-particle model.
This method is based on Heisenberg matrix mechanics, where exact
eigenstates are used, but the matrix elements of operators are the
unknown quantities. In principle the formalism gives the same results
as the Schr\"odinger equation. In practical approximations the method
leads to interesting approximate models. Especially when some
matrix elements are known experimentally, we can construct
models that are a hybrid between a phenomenological
and microscopic model.
In the case of core-particle coupling
such an approach  has been introduced by
D\"onau and Frauendorf. We study the usefulness of this approach
in a serious numerical application to the rotational nuclei
$^{155,157}$Gd,
 and extend the method to calculations of transition strengths.

Our calculations, restricted
to deformed nuclei in the current application, generalize
the particle rotor model, and we shall compare our results
to those of less microscopic calculations using this model.
The  model, first introduced by Bohr and Mottelson \cite{fp:1},
has been used for the study of both low and high spin states
of weakly and strongly deformed nuclei. For the past three decades it
has continued to provide a framework for the analysis of the
rotational band structure of odd nuclei. In spite of its approximate
nature, it has been used successfully to explain
both the structure of  deformed and transitional nuclei \cite{fp:10}.

In its most primitive version one couples a single particle
to a rotating (even-even) rigidly deformed core.
The coupling of the single particle to the core
is generally approximated by a deformed Nilsson potential.
Due to the rotation there is also a Coriolis interaction between
the total angular momentum  of the system and
the single-particle angular momentum.
In later applications one finds extensions of the original model
(we shall use calculations for $^{155,157}$Gd as references throughout
the discussion).
The  Nilsson potential can be replaced by a
more realistic deformed Woods-Saxon potential \cite{a:1},
pair correlations can be taken
into account by a quasi-particle transformation of BCS type
\cite{a:1,a:2,b:3,b:4}, and mixing between different major shells
(states with principal
quantum number $N$ differing by two) \cite{a:5,a:6,b:1,b:2,b:3,b:4}
can be included.
Finally one can allow for phonon excitations of the core
\cite{a:1,a:2,a:3,a:4},
in contrast with
the standard core-particle coupling model (CPC model) where the core is
assumed  to be a rigid rotor.

Even these extended models often suffer from an inadequacy, the so called
Coriolis attenuation problem. In order to reproduce the experimental
energy spectra, the Coriolis matrix elements have to be reduced
by up to $50\%$. This problem has been studied extensively over
the years. Many physical effects previously omitted in the model
have been invoked as the cure for this disease:
the proper treatment of the two body recoil effect
\cite{fp:6}, the inclusion of  octupole degrees of freedom
\cite{lead:1},
and the use
of an angular-momentum dependent  moment of inertia and
pairing gap \cite{fp:7}. A more complete list of all possible
suggested causes is given in Ref.~\cite{fp:8}.
It is quite conceivable that one needs to combine several of these
extensions  to eliminate this problem. As yet, however, no clear
consensus has emerged concerning the source of the attenuation problem.

In the present paper we adopt the Kerman-Klein approach, emphasizing the
steps required in the transition between a strict microscopic approach
and
one which contains compromises aimed at making applications easier to
carry out.
In effect we shall trace the
steps of  D\"{o}nau and Frauendorf in their core-quasiparticle
coupling model
\cite{fp:10,fp:11,fp:12}.
In their work a simple pairing and
quadrupole-quadrupole separable interaction Hamiltonian was considered.
Only a few application of this type exist, see Refs.~\cite{DF1,DF2}
and the review  in Ref.~\cite{fp:8}.
In our work we also introduce mixing between states with
$\Delta N = \pm 2$,
where $N$ is the principal quantum number. We believe that the pursuit
of a microscopic approach  will lead to a more complete CPC, and
give some insight into the problems of  phenomenological CPC models.
In addition the  microscopic basis allows for the straightforward
inclusion of
different  effective interactions.

This paper is organized as follows.
In Sec. 2, we discuss the basis of the model. First we describe  briefly
the basis of the Kerman-Klein method and then
summarize the D\"{o}nau-Frauendorf modifications
\cite{fp:10,fp:11,fp:12}.
We then turn to the application of a slightly extended form of the
D\"{o}nau-Frauendorf formulation to the nuclei $^{155,157}$Gd. These form
an excellent testing ground for our model since
extensive calculations have been performed for these nuclei.
There also exists a body of recent experimental data on the
transition strengths that has not yet been explored sufficiently
\cite{e:1}.
Finally, in the last section, we give a review of the other theoretical
descriptions available for  these nuclei, and compare our results with
these previous efforts.

\section{The Model}
\subsection{Kerman-Klein Method}
The Kerman-Klein method was introduced in the early sixties \cite{d:1}
to provide a rotationally invariant description of
deformed nuclei in the framework based on the equations of motion (EOM).
Among many applications, it was applied to a study of the foundations
of core-particle coupling models \cite{kk:1,f:5}.
Overall
it has been applied to various nuclear many-body problems
\cite{KleinReview} as well
as to field theories \cite{Cebl1,Cebl2}. Here
we use it as the starting point to derive and extend the
CPC model. As has been stated in the introduction, one of the
major difficulties with the CPC model is the Coriolis attenuation
problem.
We believe that the Kerman-Klein method may be a good
starting point for an investigation of this problem.
The method starts from a shell model Hamiltonian and provides
expressions for the properties of low-lying collective states of
an odd-mass nucleus
and its even-even neighbors. These expressions include the energy
eigenvalues, the single-particle
coefficients of fractional parentage (cfp)
and matrix elements of the electromagnetic operators.
In practice, Hamiltonians with only simple ingredients have been studied
(in the simplest case  single particle Hamiltonian plus
monopole pairing and quadrupole-quadrupole interactions),
and only relatively
few low-lying collective states in the even cores have been included
(in the simplest case only the ground-state rotational band).
Since the Hamiltonian must include expressions for both the
long-range part and the short-range part of the nuclear force,
we start with a Hamiltonian containing
multipole expansions involving particle-hole as well as
particle-particle contributions. This separation simplifies
the application of the EOM method.

We thus write

\begin{eqnarray}
     H
   =
     \sum_{a}h_{a} a^{\dag}_{\alpha} a_{\alpha}
  &+&
       \frac{1}{8} \sum_{abcd} \sum_{LM_{L}}
         {F_{acdb}(L)}   \,
       {B^{\dag}_{LM_{L}}(a,c)} {B_{LM_{L}}(d,b)}
   \nonumber \\
   &+&
       \frac{1}{8}
       \sum_{abcd} \sum_{LM_{L}}
            G_{abcd}(L) \,
       A^{\dag}_{LM_{L}}(a,b) A_{LM_{L}}(c,d)
. \label{eq:m1}
\end{eqnarray}
Here $h_{a}$ ($\alpha=(j_{a},m_{a})$ and $a=j_{a}$)
are the spherical single particle
energies, the operator
$B_{LM}$  is the particle-hole multipole operator,
\begin{equation}
       B^{\dag}_{LM_{L}}(a,b)
    \equiv
       \sum_{m_{a}m_{b}} s_{\beta} \;
       {C^{LM_{L}}_{\alpha\bar{\beta}}}  \;
       {a^{\dag}_{\alpha}} {a_{\beta}}
    =
       \left[
          a_{a}^{\dag} \times \tilde{a}_{b}
       \right] _{M_{L}}^{L} ,
\end{equation}
and  $A_{LM}$ is the
particle-particle multipole operator

\begin{equation}
          A^{\dag}_{LM_{L}}(a,b)
       \equiv
         \sum_{m_{a}m_{b}}
          C^{LM_{L}}_{\alpha\beta}  \;
         {a^{\dag}_{\alpha}}{a^{\dag}_{\beta}}
       =
         \left[
            a_{a}^{\dag} \times a_{b}^{\dag}
         \right] _{M_{L}}^{L}
,
\end{equation}
where $C_{\alpha\beta}^{LM}$ are the Clebsch-Gordan coefficients
and $s_{\alpha}=(-)^{j_{a}-m_{a}}$.
The coefficients $F$  are the particle-hole matrix elements

\begin{equation}
    {F_{acdb}(L)}
    \equiv
       2 \sum_{m's}{s_{\gamma}}{s_{\beta}} \;
       {C^{LM_{L}}_{\alpha\bar{\gamma}}} \,
       {C^{LM_{L}}_{\delta\bar{\beta}}}     \,
   \,  V_{\alpha\beta\gamma\delta}
,
\label{eq:f}
\end{equation}
and $G$ the particle-particle matrix elements
\begin{equation}
     G_{abcd}(L)
   \equiv
     2 \sum_{m's}
     {C^{LM_{L}}_{\alpha\beta}}    \,
     {C^{LM_{L}}_{\gamma\delta}}   \,
     V_{\alpha\beta\gamma\delta}
{}.
\label{eq:g}
\end{equation}

The task we set ourselves is to obtain equations for the states and
energies
of the odd nucleus assuming that the properties of the even nuclei
are known.
The states of the odd nucleus (particle number $N$) are designated as
$\left|\nu
 \,J\mu \right>$
where $\nu $ denotes all quantum numbers beside the angular
momentum $J$ and its projection $\mu$. The eigenstates and eigenvalues
of the neighboring even nuclei with particle numbers
($N\pm1$)
are $\left|IM\,n\,(N\pm1) \right>$ and
$E_{In}^{N\pm1}$, respectively,
where $n$ plays the same role for even nuclei as
$\nu $ does for the odd nuclei. The equations of motion (EOM)
are obtained by forming commutators between the Hamiltonian and single
fermion operators,
\begin{eqnarray}
     \left[ a_{\alpha} ,H  \right]
   =
     h_{a} a_{a}
  &+& \frac{1}{4} \sum_{bd \gamma}\sum_{LM}
     C_{\alpha \gamma}^{LM} G_{acbd}(L) a^{\dag}_{\gamma}A_{L M}(c,d)
   \nonumber \\
  &+& \frac{1}{4} \sum_{bd \gamma}\sum_{LM}
     {s_{\gamma}}C_{\alpha \bar{\gamma}}^{LM}
     F_{acdb}(L) a_{\gamma}B_{L M}(d,b)
, \label{eq:pqEOMa1}
\end{eqnarray}
\begin{eqnarray}
    \left[ a^{\dag}_{\alpha} ,H  \right]
  =
  	-h_{a} a_{a}^{\dag}
  &-& \frac{1}{4} \sum_{bd \gamma}\sum_{LM}
    C_{\alpha \gamma}^{LM} G_{acbd}(L)A^{\dag}_{L M}(c,d) a_{\gamma}
  \nonumber \\
 &-& \frac{1}{4} \sum_{bd \gamma}\sum_{LM}
    {s_{\gamma}}C_{\alpha \bar{\gamma}}^{LM} F_{acdb}(L)
   B^{\dag}_{L M}(d,b) a^{\dag}_{\gamma} .
\label{eq:pqEOMb1}
\end{eqnarray}

The matrix elements of these equations provide expressions for
the single particle coefficients of fractional parentage $U$ and $V$,
defined as
\begin{eqnarray}
     U^{\nu }_{J\mu}(\alpha; IM n) =
     \left< \nu  J\mu \mid a_{\bar{\alpha}}^{\dag} \mid IMn(N-1)
                                                               \right>
,
   \\
     V^{\nu }_{J\mu}(\alpha;IM n) =
     \left< \nu  J\mu \mid a_{\alpha} \mid IMn(N+1)
                                                   \right>
{}.
\end{eqnarray}
These are obtained by the evaluation of matrix elements of the EOM
(\ref{eq:pqEOMa1},\ref{eq:pqEOMb1})
between eigenstates $\left| \nu  J \mu \right>$  and $\left| IMn(N\pm1)
\right>$
and the use of the completeness relation (details of the derivation
can be found in \cite{d:1}).
The resulting equations
can be cast into the block matrix form,
\begin{equation}
    \left(    \begin{array}{cc}
               \epsilon + \omega^{N+1} + \Gamma^{N+1}
                             &  \Delta^{N+1} \\  & \\
       \Delta^{\dag N-1}  &  -\epsilon + \omega^{N-1} - \Gamma^{\dag N-1}
               \end{array}
    \right)
    \left(     \begin{array}{c}
                 U   \\   \\
                 V
               \end{array}
     \right)
    =
     {\cal E}_{\nu  J}
     \left(     \begin{array}{c}
                 U   \\    \\
                 V
               \end{array}
    \right) .
\end{equation}
Here $\epsilon$ are the single particle energies measured relative to
the chemical potential, $\lambda$, the elements of  $\omega$ are the
excitation energies
in the even nuclei, and $\Gamma$ and $\Delta$ are pair and
multipole fields. More specifically we have

\begin{eqnarray}
     \epsilon_{\alpha IMn,\gamma I'M'n'}
 	        & = & (\lambda_{N}  - h_{a} )\delta_{\alpha \gamma}
\delta_{II'}
 					 \delta_{MM'}\delta_{nn'}  , \\
    \lambda_{N}
		& = & \frac{1}{2} \left[ E_{0}^{N+1} - E_{0}^{N-1}
\right]
			,
                \label{eq:chemp}       \\
    \omega^{N\pm1}_{\alpha I M n, \gamma I'M'n'}
          	& =& ( E_{In}^{N\pm1} - E_{0}^{N\pm1})
			 \delta_{\alpha \gamma} \delta_{II'}
 					 \delta_{MM'}\delta_{nn'},
\end{eqnarray}
\begin{eqnarray}
       \Gamma^{N\pm1}_{\alpha IMn, \gamma I'M'n'}
		= \sum_{LM_{L}}\sum_{bd}
                 {s_{\gamma}}C_{\alpha \bar{\gamma}}^{LM_{L}}
                 F_{acdb}(L)  \hspace*{5cm}
 \nonumber \\
            \times      \left<I'M'n'(N\pm1) \mid B_{L M_{L}}(d,b)
                                    \mid IMn(N\pm1) \right>, &&
    \\
       \Delta^{N\pm1}_{\alpha IMn, \gamma I'M'n}
		= \sum_{LM_{L}}\sum_{bd}
                C_{\alpha \gamma}^{LM_{L}}
                 G_{acdb}(L) \hspace*{5cm} &&
\nonumber \\
	               \times  \left<I'M'n'(N-1) \mid A_{L M_{L}}(d,b)
                                 \mid IMn(N+1) \right>. &&
 \end{eqnarray}

In order to specify the solutions uniquely a normalization condition is
required.
{}From the anti-commutation relations of the fermion operators, one takes
the
appropriate matrix elements using a sum over intermediate states and
finds
\begin{equation}
    \frac{1}{\Omega} \sum_{\alpha \,IMn}
    \left[
         |U^{\nu }_{J\mu}(\alpha IM;n)|^2 +
	 |V^{\nu }_{J\mu}(\alpha IM; n)|^2
    \right]
        =1
,
\end{equation}
where
\begin{equation}
 	\Omega = \sum_{j_{a}} (2j_{a} +1 ).
\end{equation}

All of the above equations are still exact and represent
a formulation of Heisenberg's matrix mechanics.
The fundamental practical problem is to find a suitable `collective'
subspace
of the complete Hilbert space. We assume
that this `collective' subspace decouples
from the remaining states, at least approximately.
This requires that any matrix element of
the relevant multipole operators
between  a collective and a non-collective state is negligible.

%
\subsection{The D\"{o}nau-Frauendorf Model}
In this section we introduce the approximations made by D\"{o}nau and
Frauendorf, and use these as the practical basis for our
investigation. First we assume that we have a finite sum of
separable interactions, where each  of the terms in
(\ref{eq:f}) and (\ref{eq:g}) is of the form
\begin{eqnarray}
    F_{acdb}(L)=-2 \kappa_{L} F_{ac}(L) F_{db}(L) , \\
	G_{acdb}(L) = -2  g_{L} G_{ac}(L) G_{db}(L).
\end{eqnarray}
Here $\kappa_{L}$, $g_{L}$ are the strengths of the corresponding
multipole and pairing forces, and $F_{ac}(L)$ and $G_{ac}(L)$ are the
appropriate single particle matrix elements.
For the moment
we consider only monopole-pairing and quadrupole-quadrupole interaction
(the
inclusion of hexadecapole and the quadrupole-pairing interactions is
left
for later work). The quadrupole-quadrupole and monopole-pairing
operators are
\begin{eqnarray}
    Q_{M} = \sum_{ac}  F_{ac}(2) B_{2M} (a,c)  ,\\
	\Delta = \sum_{ac} G_{ac}(0) A_{00} (a,c).
\end{eqnarray}

co-workers have adopted
Next we identify the ``collective'' subspace appropriate for the
applications that follow.
The obvious choice is the ground
state rotational band $ \left| IM K=0 \right> $, where $K$ is the
projection of
angular momentum on the body fixed axis. The fact that this
``collective''
subspace approximately decouples from the remaining states is
equivalent to the
requirement that inter-band $BE(2)$ transitions are much smaller than
the intra-band transitions. This is generally true for deformed nuclei.
A last approximation for deformed nuclei, convenient but not necessary,
is to replace the excitation energies as well as the multipole and
pairing fields by their average values for the two neighboring
even-even nuclei.

Before we go on to the next step it is convenient to rewrite the EOM
with
reduced matrix elements,
using the Wigner-Eckart theorem. We can thus derive
equations of motion, again dropping all indices,
\begin{equation}
    \left(    \begin{array}{cc}
               \epsilon + \omega + \Gamma
                             &  \Delta \\  & \\
           \Delta  &  -\epsilon + \omega - \Gamma
               \end{array}
    \right)
    \left(     \begin{array}{c}
                 V   \\   \\
                 U
               \end{array}
     \right)
    =
     {\cal E}_{\nu  J}
     \left(     \begin{array}{c}
                 V   \\    \\
                 U
               \end{array}
    \right) ,
  \label{eq:Deom1}
\end{equation}
where $V$ and $U$ are the reduced cfp
\begin{eqnarray}
    {U_J^\lambda(1;I)}{a}
 &\equiv&
    \left<\nu  J \parallel a^{\dag}_{a} \parallel I(N-1) \right>
,  \nonumber \\
     {V_J^\lambda(1;I)}{a}
  &\equiv&
    \left<\nu  J \parallel \tilde{a}_{a} \parallel I(N+1) \right>
,
\label{eq:rcfp}
\end{eqnarray}
the self-consistent fields are represented by the expressions
\begin{equation}
    \Gamma_{aI,cI'} = -\frac{1}{2}
                     \kappa_{2} (-)^{J_{c}+I+J}
{\setlength{\arraycolsep}{2pt}
\left\{\begin{array}{ccc}
{\scriptstyle j_{a}}&{\scriptstyle {j_{c}}}&{\scriptstyle 2}\\
{\scriptstyle I'}&{\scriptstyle I}&{\scriptstyle J}
\end{array}\right\}}
                  \left< I \parallel Q_{2} \parallel I' \right>
                    F_{ac}(2),
\end{equation}
and the generalized pairing force is approximated by a constant gap
energy for all levels,
\begin{equation}
	\Delta_{aI,cI'} =
	\left< I \| \Delta \| I' \right> \delta_{ac}
\approx  \left< 0 \| \Delta \| 0 \right>
		\delta_{II'} \delta_{ac}
{}.
\end{equation}
The normalization condition
becomes
\begin{equation}
    \frac{1}{\Omega} \sum_{a \,I}
    \left[
         |{U_J^\lambda(1;I')}{a}|^2 + |{V_J^\lambda(1;I')}{a}|^2
    \right]
        = 1
. \label{eq:ortho}
\end{equation}
Here the inclusion of only the collective subspace is somewhat more
questionable than for the EOM, but there appears to be no simple
alternative.

The final step in the derivation is the replacement of
the self consistent fields by phenomenological inputs.
In this context $\left< I \| Q \| I' \right>$ and $\Delta$ are the
quadrupole matrix elements
of the neighboring even nuclei and $\Delta$ their gap energies.
The quadrupole fields can either be extracted from
experiment
or calculated using a phenomenological model such as the Bohr-Mottelson
model
\cite{fp:1},
\begin{equation}
	\left< I' \| Q \| I \right>
	=  q_{0} \, \sqrt{ \frac{5} {16 \pi}} \,
		 \sqrt{(2I+1)(2I'+1)} \,
{\setlength{\arraycolsep}{2pt}\left(\begin{array}{ccc}
{\scriptstyle I}&{\scriptstyle 0}&{\scriptstyle I'}\\
{\scriptstyle 0}&{\scriptstyle 2}&{\scriptstyle 0}\end{array}\right)}
 , \label{eq:b&m}
\end{equation}
where $q_{0}$ is a phenomenological input which can be obtained from
experimental values.

Having described the assumptions involved in the derivation
of the model, we turn to the task of solving the resulting equations.
The main difficulty in solving those
equations is that the
set of solutions (\ref{eq:Deom1})
is over-complete by a factor of two. This is a consequence of the
fact that the basis states $a^{\dag} \left| I(N-1) \right>$ and
$a \left| I(N+1) \right>$ form an over-complete and non-orthogonal set
as in the standard BCS model.
Half of the states found by solving (\ref{eq:Deom1})
 are spurious, and these have to
be identified.
To remove the spurious solutions one can proceed in the following way
\cite{fp:11,fp:12}: First build the
density matrix

\begin{eqnarray}
     D_{aa'}(II') = \sum_{\nu J} \hspace{4in} \nonumber \\
       \left[
      \begin{array}{cc}
        \left< I(N-1) \| a_{a} \| \nu  J \right> \left< \nu  J
	\| a^{\dag}_{a'} \| I'(N-1)\right>  &
        \left< I(N+1) \| a^{\dag}_{a} \| \nu  J \right> \left< \nu  J \|
	a^{\dag}_{a'} \| I'(N-1) \right> \\
                                                             \\
        \left< I(N-1) \| a_{a} \| \nu  J \right> \left< \nu  J \|
	a_{a'} \| I'(N+1) \right>  &
        \left< I(N+1) \| a^{\dag}_{a} \| \nu  J \right> \left< \nu  J \|
	a_{a'} \| I'(N+1) \right>
      \end{array}   \right]
,
\end{eqnarray}
which can also be expressed
in terms of the reduced cfp $U$ and $V$,
\begin{equation}
     D_{aa'}(II') = \sum_{J} D_{aa'}^{J}(II')
      = \sum_{J\nu }
       \left[ \begin{array}{cc}
         {U_J^\lambda(1;I)}{\alpha} {U_J^\lambda(1;I')}{\alpha'}  &
{U_J^\lambda(1;I)}{\alpha} {V_J^\lambda(1;I')}{\alpha'} \\  \\
        {V_J^\lambda(1;I)}{\alpha} {U_J^\lambda(1;I')}{\alpha'}  &
{V_J^\lambda(1;I)}{\alpha} {V_J^\lambda(1;I')}{\alpha'}
            \end{array} \right]
\label{eq:dm}
{}.
\end{equation}
This density matrix has two important properties; it commutes with
Hamiltonian
\begin{equation}
 \left[ H, D \right ]_{J}=0
, \label{eq:commutes}
\end{equation}
where $H$ is the matrix Hamiltonian of Eq. (\ref{eq:Deom1})
and the scaled quantity $\Omega \,$D acts like a projection operator
\begin{equation}
(\Omega D)^{2}= \Omega D
. \label{eq:proj}
\end{equation}
(Here (\ref{eq:commutes}) follows from the EOM and (\ref{eq:proj})
expresses
the
ortho-normalization conditions that follow from (\ref{eq:Deom1}) and
(\ref{eq:ortho}).)

Therefore, $D$ has eigenvalues $1/\Omega$ or 0. The eigenstates of the
Hamiltonian
which are also eigenstates of $D$ with eigenvalue
$1/\Omega$ describe the real quasiparticle excitations in the odd-$N$
nucleus, whereas the eigenstates with eigenvalue 0
characterize the spurious states.

We next decompose the Hamiltonian as,

\begin{equation}
   H=H_{qp} + H_{c}    ,
\end{equation}
where
\begin{equation}
   H_{qp} = \left (
          \begin{array}{cc}
           -\epsilon-\Gamma  & \Delta \\ \Delta & \epsilon+\Gamma
          \end{array} \right ) , \hspace*{1cm}
       H_{c} = \left( \begin{array}{cc} \omega_{c} & 0 \\ 0& \omega_{c}
                       \end{array} \right)
{}.
\end{equation}
$H_{qp}$ is interpreted as a generalized quasi-particle
Hamiltonian and $H_{c}$ as the core Hamiltonian.
$H$ is separated into two parts because $H_{qp}$ is antisymmetric
with respect to particle-hole conjugation ($a_{a}^{\dag}
\rightarrow a_{a}$
), and therefore, as in the usual BCS
theory \cite{fp:15}, the solutions of this part of the
Hamiltonian are divided into two sets with sign-reversed energies;
the negative energy solutions are the spurious states.
This decomposition gives a good starting point for a numerical technique
 to remove
the spurious  states. First, we turn off the core Hamiltonian $H_{c}$.
We solve the eigenvalue problem,
$
    H_{qp} \Psi^{0} = {\cal E}^{0} \Psi^{0}
$, and find the solutions $\Psi^{0}_{\pm}$ and $\pm{\cal E}^{0}$
where
\begin{displaymath}
    \Psi^{0}= \left( \begin{array}{c} V^{0} \\ U^{0} \end{array}
\right)
{}.
\end{displaymath}
{}From the physical states
 $\Psi_{+}^{0}$ we can construct the density matrix
  $D^{0}$ as in (\ref{eq:dm}).
Then the core Hamiltonian is gradually turned on,
\begin{equation}
   H(\gamma)=H_{qp} + \gamma H_{c},
\label{eq:adiab}
\end{equation}
where $\gamma$ is a scaling parameter. The next step is to diagonalize
$H(\gamma)$ and find the
new solutions, $\Psi(\gamma)$. If the change of $\gamma$
is not too big, the quantity $\Psi^{\dag}(\gamma) D^{0} \Psi(\gamma)$
  will either be
close to 0 or to $1/\Omega$.
When the value of $\Psi^{\dag}(\gamma) D^{0} \Psi(\gamma)$  is
close to $1/\Omega$ it is a real state and when it is close to zero it
is a spurious state. Now we can
construct the new density matrix from the physical states
$\Psi_{+}(\gamma)$ which we
call $D^{1}$. Then $\gamma$ is incrementally increased and this
process is repeated until $\gamma=1$.
In practice this method works very well since
the change in the eigenvalues of $D$ never exceeds $5\%$, even with
only 15 incremental steps.

When $\gamma \rightarrow 0$, which is the limit that the moment of
inertia
$ {\cal I}$ goes to infinity, the core excitations
vanish (adiabatic limit). In this limit, $H_{qp}$
is equivalent to the usual Nilsson plus pairing
Hamiltonian. To demonstrate this, we first diagonalize the particle-hole
part ($\epsilon+\Gamma$) of $H_{qp}$. The resulting eigenvalues
$e_{k}-\lambda$
are characterized by the index $k$. For each different $k$,
$\Gamma$ defines a different fixed
quadrupole field or, in other words, an instantaneous orientation of the
core. The index $k$ can be related to $K$, the
projection of the angular momentum on the body fixed frame, by a method
described below. Finally,
the solutions of the $H_{qp}$, including pairing, are
\begin{equation}
     {\cal E}^{\pm}_{K}= \pm \sqrt{(e_{k}-\lambda)^{2}+\Delta^{2}}
{}.
\end{equation}

The remaining problem is to find the correspondence between the
values of $K$
and those of $k$.
This is
not straightforward because H$_{qp}$ is an angular momentum conserving
quantity; therefore $K$ can not be obviously associated with the
eigenvalues. We use the fact that among states with different $J$,
states with the
same value of $K$ have identical eigenvalues. We can
extract a unique state of maximal $K$ value in the following way:
Compare the $J+1/2$ doubly degenerate solutions for angular momentum $J$
with the $J-1/2$ doubly degenerate solutions for $J-1$.
Since the maximum values of $K$ are
$J$ and $J-1$, respectively, the additional pair of
solutions for angular momentum $J$ must have $|K|=J$.
We then proceed with $J-2$, and so on.

\subsection{Electromagnetic Transitions}
The electromagnetic multipole operators are one-body matrix
elements of the form
\begin{equation}
  T_{LM} = \sum_{\alpha\gamma}^{\pi \nu} e_{\alpha\gamma}
t_{\alpha\gamma} a^{\dag}_{\alpha}a_{\gamma}
  ,
	\label{eq:be1.1}
\end{equation}
where $e_{\alpha\gamma}$ is either the effective charge or the effective
$g$ factor,
and $t_{\alpha\gamma}$ are the single-particle matrix elements of the
transition operator.

It is convenient to decompose the electromagnetic observables into core
and
particle contribution, as in the conventional CPC.
To do this we must express the matrix elements
of the transition multipole operator of the odd nucleus,
$\left< J' \mu' \nu' | T_{LM} | J \mu \nu \right >$, in terms of
the matrix elements of the even nuclei (which are
known from experiment) and the known eigenfunctions $U$ and $V$.
We use the decomposition
of the eigenstates of the odd nucleus \cite{fp:10,fp:11},
\begin{equation}
	  	\left| J \mu \nu \right>
		= \frac{1}{\Omega}
 \sum_{IM\alpha} \left[
		U_{J\mu}^{\nu}(\alpha;IM)\, a_{\alpha}^{\dag} \,
\left| IM \right>
	+       V_{J\mu}^{\nu}(\alpha;IM) \,a_{\alpha}
		\, \left| \widetilde{IM} \right>\right]
,
\end{equation}
where
\begin{eqnarray}
	\left| IM \right> &=& \left| IM (N-1) \right>  , \nonumber \\
	\left| \widetilde{IM} \right>  &=&  \left| IM (N+1) \right>.
\end{eqnarray}
We then rearrange the order of the single particle operators $a$ and
$a^{\dag}$
and the collective operator $T$
using their commutation relations
and utilize sums over intermediate states of the
even nuclei such that operator $T$ stands between states of the even n
uclei
and the single particle operators occur between an even nucleus  state
and an odd nucleus state. Finally we have
\begin{eqnarray}
   \left< J' \mu' \nu' | T_{Lm} | J \mu \nu \right > &  = &
	 	 \frac{1}{\Omega}\sum_{IM\alpha} \sum_{I'M'}
		\left[	U_{J\mu}^{\nu}(\alpha;IM)
			 U_{J'\mu'}^{\nu'}(\alpha;I'M')
		\left< I'M' \mid T_{Lm} \mid \widetilde{IM} \right>
\right.
		\label{eq:be1.sf}			  \\
		&&  \hspace*{0.2in} \left.
	+  V_{J\mu}^{\nu}(\alpha;IM)
		 V_{J'\mu'}^{\nu'}(\alpha;I'M') \left< I'M' \mid
			T_{Lm} \mid \widetilde{IM} \right>  \right]
	 \nonumber \\
		&+& \frac{1}{\Omega} \sum_{IM\alpha}
			 \sum_{\gamma} e_{\alpha \gamma} t_{\alpha\gamma}
  		\left[ U_{J\mu}^{\nu}(\alpha;IM)
			 U_{J'\mu'}^{\nu'}(\gamma;IM)
		+ V_{J\mu}^{\nu}(\alpha;IM)
				 V_{J'\mu'}^{\nu'}(\gamma;IM) \right].
	\nonumber
\end{eqnarray}
Equation (\ref{eq:be1.sf}) clearly exhibits a separation into particle
and core contributions.
With the definition of the reduced matrix elements (\ref{eq:rcfp})
we ultimately obtain
\begin{eqnarray}
\left< J'\nu' \| T_{L}  \| J \nu\right> &=& \frac{1}{\Omega} \sum_{a II'}
				   \left<I\| T_{L} \| I'\right>
		(-)^{j_{a}+J+I'} \,
{\setlength{\arraycolsep}{2pt}\left\{\begin{array}{ccc}
{\scriptstyle I'}&{\scriptstyle I}&{\scriptstyle 2}\\
{\scriptstyle J}&{\scriptstyle J'}&{\scriptstyle j_{a}}
\end{array}\right\}}
	\left[ U_{J}^{\nu}(a;I) U_{J'}^{\nu'}(a;I')
			 + V_{J}^{\nu}(a;I) V_{J'}^{\nu'}(a;I')
			\right]  \nonumber \\
	&+&  \frac{1}{\Omega} \sum_{Iac} (-)^{j_{a}+J'+I} e_{ac} t_{ac}
{\setlength{\arraycolsep}{2pt}
\left\{\begin{array}{ccc}
{\scriptstyle J}&{\scriptstyle J'}&{\scriptstyle 2}\\
{\scriptstyle j_{c}}&{\scriptstyle {j_{a}}}&{\scriptstyle I}
\end{array}\right\}}
		 \left[ U_{J}(a;I) U_{J'}^{\nu'}(c;I)
			+ V_{J}(a;I) V_{J'}^{\nu'}(c;I)
			\right]
 ,
\label{eq:be2}
\end{eqnarray}
	where we assume that the matrix elements of $T_{L}$ do not
depend on $N$,
\begin{equation}
	\left< I \| T_{L}  \| I' \right > = \left< \tilde{I} \| T_{L}
\| \tilde{I'} \right>
{}.
\end{equation}

For the electric quadrupole transitions studied in this paper,
the matrix elements
$\left< I \| T_{L}  \| I' \right>$
become the matrix elements of the quadrupole operator
$\left< I \| Q  \| I' \right>$.
The single particle matrix element $t_{ac}(L)$ are
\begin{equation}
    F_{ac} (L) = \left<a \parallel r^{2} Y^{2} \parallel c \right>.
\label{eq:quadsp}
\end{equation}
For these electric transitions we have only one parameter to fix, the
effective charge. As we will see later this is a minor problem since
the single particle contribution is much smaller than the core
contribution.
The quadrupole matrix elements of the core can be either derived
from the Bohr-Mottelson model as in (\ref{eq:b&m})
or extracted from the experimental results.

For the magnetic ($M1$) transitions the matrix elements of the  operator
is
\begin{equation}
\left< I \| T_{L}  \| I' \right> = \left< I \| M  \| I \right>  \
delta_{II'},
\end{equation}
and the single particle matrix elements $t_{ac}$ become
\cite{fp:15}
\begin{eqnarray}
   t_{ac}= G_{ac}& =& \left< j_{a} \parallel M \parallel j_{c} \right>
         \nonumber    \\
          & = &
	  \left<n_{a}l_{a} j_{a} \mid  n_{c} l_{c} j_{c} \right>
           \sqrt{\frac{3\,(2j_{a}+1)\,(2j_{c}+1)}{4\pi}}
         (-1)^{j_{c}-1/2}
      \left(
         \begin{array}{ccc}
          j_{a} & 1 & j_{c} \\
          -1/2 & 0 &1/2
         \end{array}
      \right) \nonumber \\
	& & \hspace{2in} \times (1-k) \left[ \frac{1}{2}g_{s}
-g_{l}(1+\frac{k}{2})\right]
, \label{eq:thm1}
\end{eqnarray}
with
\begin{equation}
	k = (j_{a} + \frac{1}{2})(-1)^{(j_{a} +l_{a} +1/2)} +
	 (j_{c} + \frac{1}{2})(-1)^{(j_{c} +l_{c} +1/2)}
{}.
\end{equation}

\section{Calculations for $^{155,157}$Gd}
\begin{figure}
\centerline{\epsfysize=8cm \epsffile{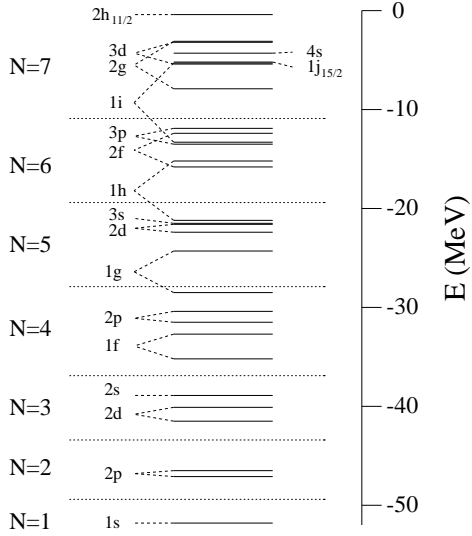}}
\caption{ \label{fig:ws1}
Single particle eigenvalues for Woods-Saxon potential . }
\end{figure}

We decided to concentrate the calculations on the nuclei $^{155,157}$Gd,
since recent detailed studies of the transitions \cite{e:1} seem to
require explanation. The even cores $^{154,156,158}$Gd are well deformed
and have almost rigid rotational bands.
Therefore, the
quadrupole field $\Gamma$ can be estimated using the Bohr-Mottelson model
as
\begin{equation}
     \Gamma_{aI,cI'} = -\frac{22.21}{2}  \, \kappa \beta A^{2/3}
(-)^{j_{c}+I+J}
{\setlength{\arraycolsep}{2pt}
\left\{\begin{array}{ccc}
{\scriptstyle j_{a}}&{\scriptstyle {j_{a}}}&{\scriptstyle 2}\\
{\scriptstyle I}&{\scriptstyle I'}&{\scriptstyle J}
\end{array}\right\}}
               \sqrt{(2I+1)(2I'+1)}
{\setlength{\arraycolsep}{2pt}\left(\begin{array}{ccc}
{\scriptstyle I}&{\scriptstyle 0}&{\scriptstyle I'}\\
{\scriptstyle 0}&{\scriptstyle 2}&{\scriptstyle 0}\end{array}\right)}
               q_{ac}
,
\end{equation}
where $q_{ac}$ are the single particle quadrupole matrix elements
(labeled
before by $F_{ac}(2)$)
in fm. The factor 22.21 comes from the transformation from
the intrinsic quadrupole moment to the deformation parameter $\beta$.
The core energies are
$\omega_{I}= \frac{I(I+1)}{ 2{\cal I}}$, where for the moments of
inertia the
values ${\cal I}^{155} = 0.0122 \,$MeV , ${\cal I}^{157} = 0.0124 \,$MeV
are chosen. For the pairing potential
we used the value
$\Delta = \frac{127}{A}$ MeV. The chemical potential $\lambda^{N}$
was calculated according to (\ref{eq:chemp}) from the difference between
the ground state energies of the cores. The values found are
$\lambda^{155}=-7.4845 \, $MeV and
$\lambda^{157}=-7.1488 \, $MeV. The experimental values for the
quadrupole
deformations of
the cores are \cite{fp:17}

\begin{displaymath}
\begin{array}{cc}
  ^{154}\,\rm{Gd}   :& \beta_{20}=0.3294,\\
   ^{156}\, \rm{Gd}   :& \beta_{20}=0.3378,\\
  ^{158}\, \rm{Gd}  :&  \beta_{20}=0.3484.
\end{array}
\end{displaymath}
For the odd nuclei we used the averages of the two neighboring even nuclei,
$\beta_{20}^{155} = 0.329$,
 $\beta_{20}^{157} = 0.3431$.
The strength of the quadrupole force $\kappa$ was taken as a free
parameter.
The single particle wave functions were found from a Woods-Saxon
potential
\cite{fp:15},

\begin{equation}
V^{ws}(r) = \frac{-V_{0}}{ 1+\exp \left( \frac{r-R_{0}}{a} \right)},
\end{equation}
including spin-orbit interaction with parameters
$a=0.65$, $R_{0}=1.25A^{1/3}$ and $V_{0} = 51/\left(1-\frac{2}{3}
\frac{N_{p}-Z}{A}\right)$ MeV.
The single particle quadrupole matrix elements $q_{ac}$ take values
according to the standard formula \cite{fp:15}
\begin{eqnarray}
    q_{ac}& =& \left< j_{a} \parallel Q \parallel j_{c} \right>
             \\
          & = &
    \left<n_{a}l_{a} j_{a} \mid r^{2} \mid n_{c} l_{c} j_{c} \right>
         \sqrt{\frac{5\,(2j_{a}+1)\,(2j_{c}+1)}{4\pi}}
         (-1)^{j_{c}-1/2}
      {\setlength{\arraycolsep}{2pt}\left(
         \begin{array}{ccc}
      {\scriptstyle j_{a}} & {\scriptstyle 2} & {\scriptstyle j_{c}} \\
       {\scriptstyle -1/2} & {\scriptstyle 0} &{\scriptstyle 1/2}
         \end{array}
      \right)  }
{}.
\end{eqnarray}
The resulting single-particle levels are shown in Fig.~\ref{fig:ws1}.

\begin{figure}
\centerline{\epsfysize=8cm \epsffile{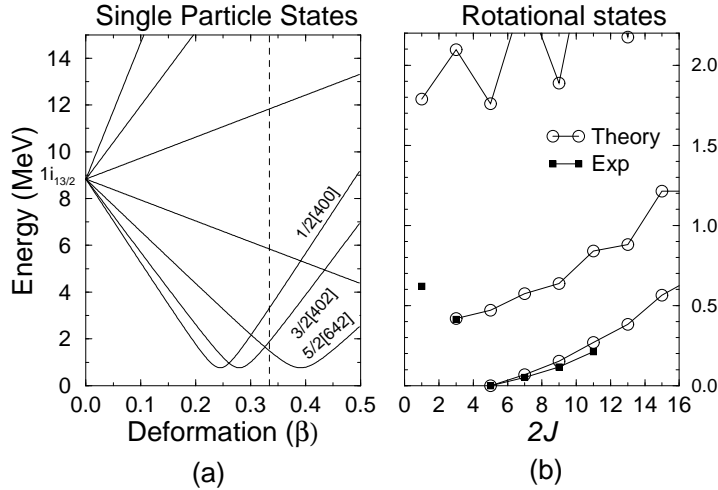}}
\caption{ \label{fig:pos1}
Energy levels for the positive parity states of $^{157}$Gd.
Only single-particle states from $N=6$ major shell are included.
The $5/2[642]$ is below the $3/2[402]$ and $1/2[400]$ at deformations
larger
than 0.33. The quadrupole-quadrupole strength is set at
$\kappa= 0.311~{\rm MeV/fm}^2$. The black squares represent the
experimental	values. }
\end{figure}

First we apply the method to positive parity states of $^{157}$Gd.
The experimental results identify  the $5/2^{+}[642]$ band as the
lowest positive parity band, followed by the $3/2^{+}[402]$ and
$1/2^{+}[400]$
bands at excitation energies $\sim0.4$ MeV and $\sim0.6$ MeV r
espectively.
Figure (\ref{fig:pos1}a) shows the solution in
the adiabatic limit ($\gamma =0$ in Eq. (\ref{eq:adiab}))
as a function of deformation using only the $1i_{13/2}$
quasi-particle state. As stated
before, this should be equivalent to the Nilsson  diagram
(including pairing).
At zero deformation we have a degeneracy because of the spherical
symmetry. As the deformation is increased the spherical symmetry is
broken
and the degeneracy is lifted. As a result there are 14 states each
 one with different $K$, taking values from $j$ to $-j$. Because of the
signature symmetry, states with $K$ having opposite signs are
degenerate.
For the case of prolate
deformation, quasi-particle states with smaller $K$ are lower in energy.
In the case of
quasi-hole states the opposite is true.
We found that for
quadrupole-quadrupole strength $\kappa=0.31~{\rm MeV/fm}^2$
the band-head energies are
in best agreement with experiment. The $5/2^{+}[642]$ state is below
the $3/2^{+}[402]$ by about $0.4$ MeV, but the $1/2^{+}[400]$ is off
 by $1$ MeV.
In Fig.~\ref{fig:pos1}b we show the result when the core excitations were
turned on. The lowest band $5/2^{+}[642]$ band reproduces the
 experimental
band and to the very limited extend that data is available, so does the
$3/2^{+}[402]$ band.
The $1/2^{+}[400]$ band is off
by 1 MeV.

For better results we include more single particle and single hole
states ;
two more major shells were included, the $N=5$ and $N=7$
(see Fig.~\ref{fig:ws1}).
For $\kappa = 0.23~{\rm MeV/fm}^2$ the band-head energies are in good
agreement with experimental values.
This change in the value of $\kappa$ was to be expected, since we are
using an effective interaction. It is therefore natural to
need different strengths
for the interaction at different dimensions of the
included single-particle space.
The zig-zag shape of the $3/2^{+}[402]$ and $1/2^{+}[400]$ rotational
bands
reveals that the Coriolis interaction is strong compared to
single particle excitations (quadrupole-quadrupole interaction)
such that it creates the staggering effect. Since there
are no experimental data for those bands, we cannot draw any firm
conclusions concerning this behavior. The fact that the $1/2^{+}[400]$
band-head is still off by 0.3 MeV may imply that a more sophisticated
interaction is
needed to fully describe the structure of this band.
\begin{figure}
\centerline{\epsfysize=8cm \epsffile{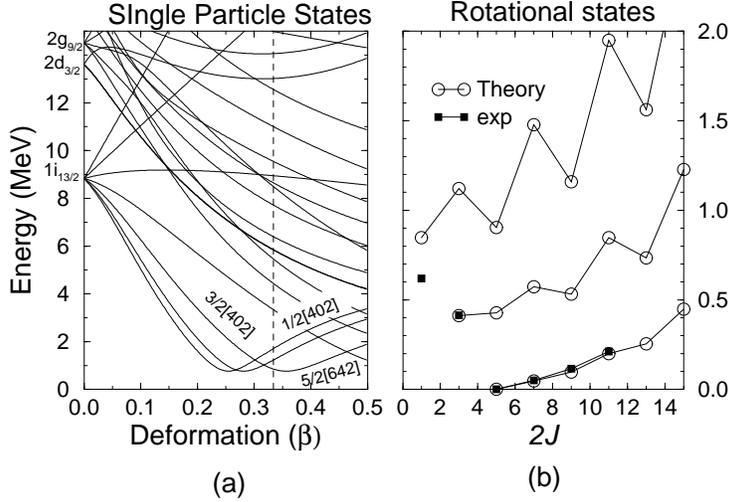}}
  \caption{ \label{fig:pos2}
The positive parity energy levels of $ ^{157}$Gd.
The circles correspond to the theoretical predictions and
the squares to the experimental values. The strength of
quadrupole-quadrupole
interaction is $\kappa = 0.23~{\rm MeV/fm}^2$. Here we include states
from 3 major shells.}
\end{figure}

\begin{figure}
\centerline{\epsfysize=8cm \epsffile{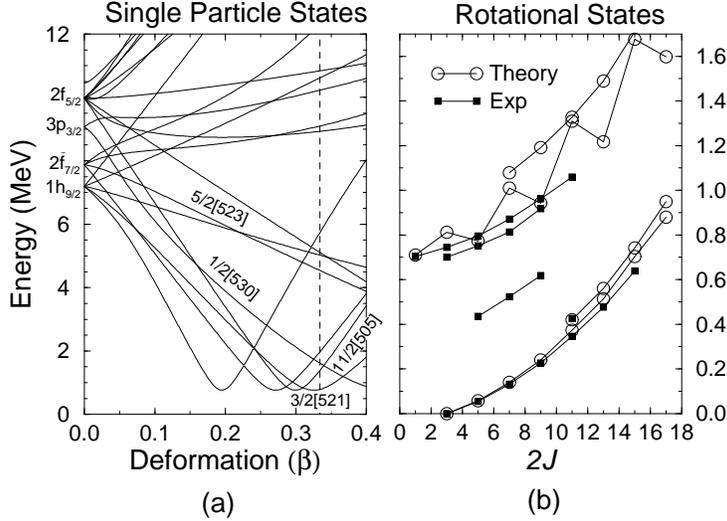}}
\caption{ \label{fig:neg1}
Negative parity states of $^{157}$Gd. The squares correspond to
experimental
values. Only states from the $N=6$  major shell are included.
The strength
of the quadrupole force  $\kappa$ is set to $0.201~{\rm MeV/fm}^2$.}
\end{figure}

The next step is the application of the method to the negative
parity states where more data is available.
The experimental situation can be summarized as follows:
The
${3/2^-\left [5 2 1 \right ]}$
is the ground state band. Then follows the ${11/2^-
\left [5 0 5 \right ]}$ hole band
and the ${5/2^-\left [5 2 3 \right ]}$ particle band.
The ${3/2^-\left [5 3 2 \right ]}$ band and the ${1/2^-
\left [5 3 0 \right ]}$ band, which are not
certain, are higher in energy and almost degenerate.
If we only include the $N=6$ major shell, the
strength $\kappa$ should be set at a value of about $0.2~{\rm MeV/fm}^2$
for the $3/2^{-}[521]$ band-head to be the ground state, which
agrees with all previous calculations \cite{f:7,f:9}.
Figure \ref{fig:neg1}a shows our ``Nilsson'' diagram.
Again particle states  with lowest $K$ have lowest energy.
The states deriving from the hole
states obey the opposite rule; the lowest state has the highest $K$
value.
In addition to this, we note that quasi-particle and quasi-hole states
originating from different single particle orbitals
interact weakly with
each other. This leads to narrow avoided crossing between states
with the same $K$ value.
States with different principle quantum number $N$ also interact
weakly. This is because the matrix elements $\left< r^{2} \right>$
for states having
$\Delta N $ not equal to zero is smaller by a factor of the order of
10 as
compared to states having the same $N$.
Quasi-particle (or quasi-hole) states having angular momentum differing
by four units and having the same $K$ couple strongly.
For
$\Delta j$ equal to 2 or 0 their interaction is weak.
This is a consequence of the relevant geometric
factors (three and six-j symbols). For example, the quasi-particle
states of $f_{5/2}$ repel the $h_{9/2}$ states with the same $K$; as
a result the band-head energies are quite sensitive to the
position of the  single-particle energy of the $f_{5/2}$ orbital.
Because of the approximate
nature of the Woods-Saxon potential, some adjustments to
the single particle energy levels will be necessary for better
results.
The $11/2^{-}[505]$ single hole state
deriving from $1h_{11/2}$ does not interact with any other state.
We can easily force the $11/2^{-}[505]$ band-head to be 0.4 MeV above
the ground state by lowering the $1h_{11/2}$ single-hole energy level by
1.5 MeV.
Figure \ref{fig:neg1}b shows the calculated band structure compared to
the experimental
values. The ground state band $3/2^{-}[521]$ and  $11/2^{-}[505]$
band-head are reproduced very accurately.
The $1/2^{-}[530]$ band-head is in good agreement with the experimental
value but the rotational structure is distorted. This is due to the
strong
Coriolis interaction.
The $5/2^{-}[523]$ and  $3/2^{-}[532]$ band-heads ares way off ($\sim$5
MeV and
$\sim$10 MeV respectively). At this point we can
no longer improve the prediction by small adjustments of the
single-particle
or single-hole energies.

In an attempt to improve
agreement with experiment we include all single particle and hole l
evels from
from the $N=4$ major shell as well as intruder states from N=5 and
N=7 shells.
As was to be expected, the strength of the interaction must be adjusted
because of the change in the size of the single
particle space.
For best results the
quadrupole-quadrupole strength was found to be
$\kappa=0.397~{\rm MeV/fm}^2$.
Though it is clear why $\kappa$ is different for
different dimensions, it is not clear
why $\kappa$ is not the same for
positive and negative parity states. For one major shell the positive
parity states require $\kappa=0.31~{\rm MeV/fm}^2$ whereas for the
negative parity states
$\kappa=0.20~{\rm MeV/fm}^2$. For three major shells the positive
parity states require
$\kappa=0.2~{\rm MeV/fm}^23$, whereas for the negative parity states
$\kappa=0.39~{\rm MeV/fm}^2$.
For the moment, this remains a puzzle.

\begin{figure}
\centerline{\epsfysize=8cm \epsffile{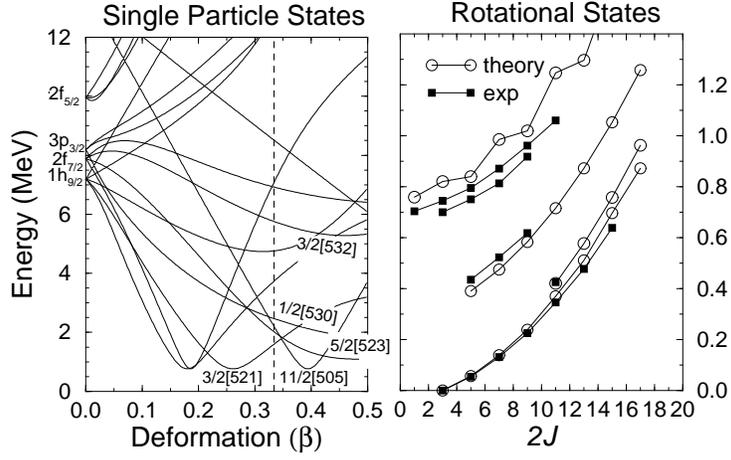}}
\caption{
\label{fig:neg2}Negative parity energy levels $^{157}$Gd.
            Single particle and hole states from three major shells
	are included. Squares correspond
 to the experimental values and circles to theoretical. The strength
$\kappa$
of the quadrupole force is $0.397~{\rm MeV/fm}^2$.}
\end{figure}

The results for this more complete calculation are shown
in Fig.~\ref{fig:neg2} and
are more satisfactory than those for one shell.
The $5/2^{-}[523]$ band-head is at the right energy and the same
is true for the
$1/2^{-}[530]$ state. The $11/2^{-}[505]$ state is at the right position
without any change to the single-particle energy
spherical energy found by solving the Woods-Saxon potential.
Only the $3/2^{-}[532]$ is off
by $\sim 1$ MeV.
The only problem is that the structure of the rotational
band $1/2^-[530]$ is slightly distorted. This again may indicate that
higher
order interactions are needed to describe states of higher energy.
The results are very satisfactory keeping in mind that very few free
parameters were used. The most important result is  that we can
reproduce the rotational structure
with accuracy better than any previous work,
without using any attenuation
or any other forms of interaction. At the same time the inclusion of the
latter remains a relatively straightforward possibility

\begin{figure}
\centerline{\epsfysize=8cm \epsffile{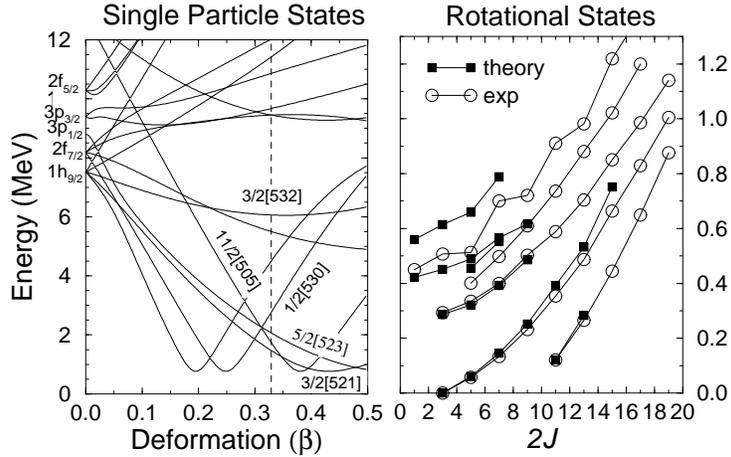}}
\caption{
\label{fig:155neg2}Negative parity energy levels for $^{155}$Gd.
            Single particle and hole states from three major shells
	are included. Squares correspond
 to the experimental values and circles to the theoretical calculations.
 The strength $\kappa$ of the quadrupole force is
$0.377~{\rm MeV/fm}^2$.}
\end{figure}

In Fig.~\ref{fig:155neg2} we show a similar calculation for
the negative parity levels of $^{155}$Gd.
Single-particle states from the $N=4$, $N=6$ as well as intruder states
from the $N=5$
and $N=7$ shells were included. The experimental structure of the bands
is
very similar to those of $^{157}$Gd with the slight change in the
band-head
energies. The quadrupole-quadrupole strength was found to be
$\kappa=0.377~{\rm MeV/fm}^2$, which is comparable to the one used for
the negative
parity states of $^{157}$Gd.
The $5/2^{-}[523]$ and the  $11/2^{-}[505]$
band-heads are reproduced well. The same is true for  $1/2^{-}[530]$
but the rotational structure of the band is distorted.
As for $^{157}$Gd, the  $3/2^{-}[532]$
is off by $\sim 1.5$ MeV. The results are as good as for $^{157}$Gd
and the
same conclusions as above apply for this nucleus.


\begin{figure}
\centerline{\epsfysize=8cm \epsffile{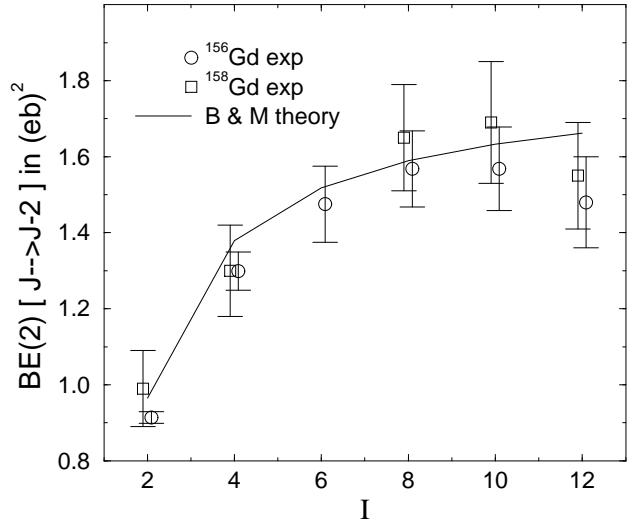}}
    \caption{ \label{fig:B&M}
The experimental values of $BE(2)$'s in two of the even-even Gd
isotopes compared
to the values calculated using Bohr-Mottelson model. }
\end{figure}

\begin{figure}
\centerline{\epsfysize=8cm \epsffile{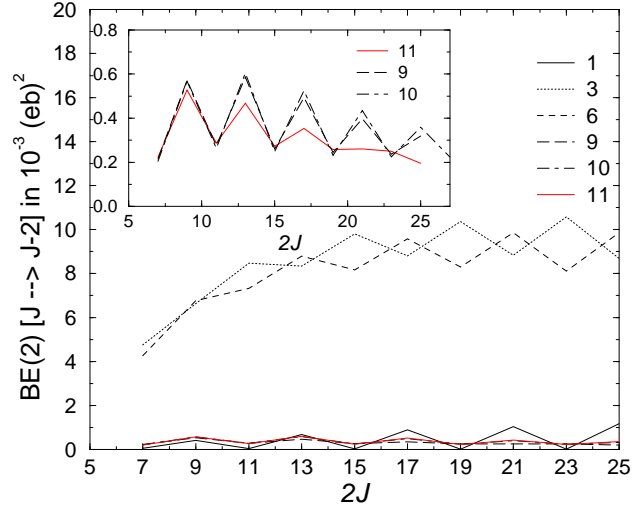}}
    \caption{ \label{fig:con}
            The single particle contribution to the $BE(2)$ $^{157}$Gd
for different numbers of single particle states, using
$e_{{\rm eff}}=1$.}
\end{figure}

\begin{figure}
\centerline{\epsfysize=8cm \epsffile{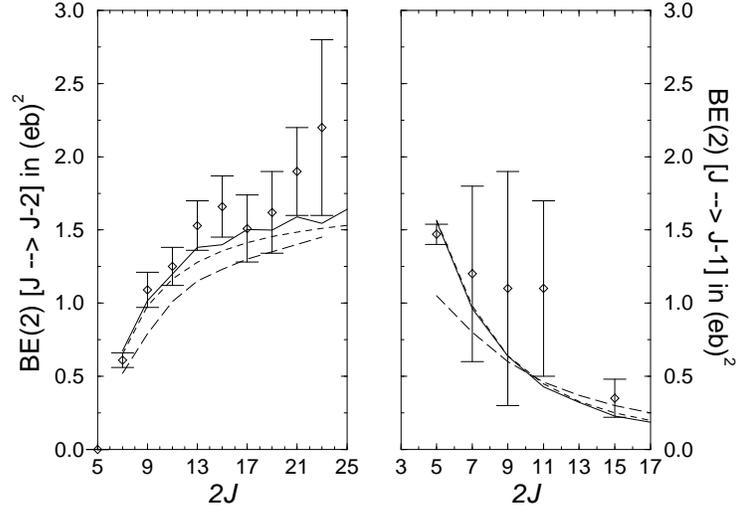}}
    \caption{ \label{fig:f1}
           Comparison of our results with other models for $^{157}$Gd.
The formulas for the Particle-Rotor Model were taken
from Ref. \protect{\cite{fp:1}}, for the Cranking model
from Ref. \protect{\cite{e:1}}. The points with error bars are the
experimental data \protect{\cite{e:1}}, the short-dashed line is the
Particle-Rotor model, the long-dashed line is the Cranking model and the
solid line is the present work.}
\end{figure}

\begin{figure}
\centerline{\epsfysize=8cm \epsffile{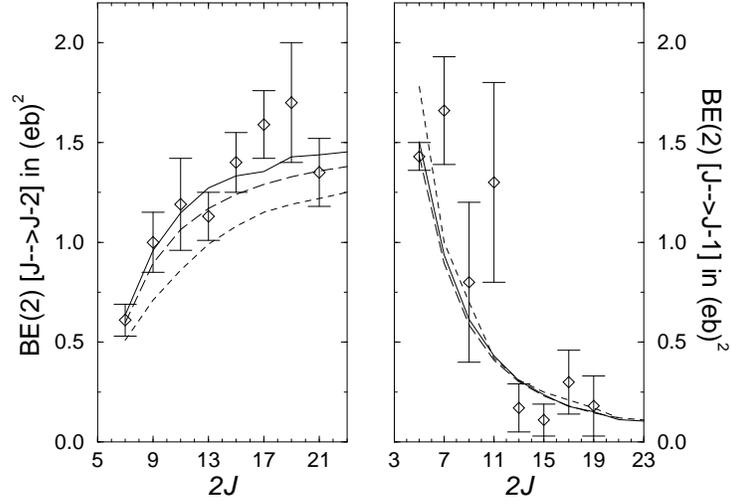}}
    \caption{ \label{fig:f2}
          Comparison of our results with other models $^{155}$Gd.
See Fig.~\protect{\ref{fig:f1}} .}
\end{figure}

We have also calculated the electromagnetic transitions
and compared them to experiment.
In Fig.~\ref{fig:B&M} we show the experimental
$BE(2)$'s for $^{156}$Gd and  $^{158}$Gd
compared to the theoretical calculations using
(\ref{eq:b&m}).
The theoretical calculation represents the average of $^{156}$Gd and
$^{158}$Gd. As can be seen from the figure the model reproduces the
experiment
reasonably well with some evidence of deviations at the
largest angular momentum. For odd nucleus states having
angular momentum $J <  19/2$, the contribution of states in the
even cores with angular momentum $I > 12$ will be
small. In the case of $^{154}$Gd and $^{156}$Gd we have a similar
situation.
Therefore, we can use either the theoretical calculations or
the experimental values.
Various aspects of the $BE(2)$ for transitions
from $J$ to $J-2$ and from $J$ to $J-1$ are illustrated in
Figs. \ref{fig:con}-\ref{fig:f2}.
The experimental data are taken from \cite{e:1}. The salient fact is
(as was expected) that the core contribution is much larger than
the single particle contribution (compare the scales in
Fig.~\ref{fig:con}
with those in Fig.~\ref{fig:f1}).
We see that the numbers of single-particle
states included only affect the single-particle contribution
but not the core contribution.
This shows that we are close to the extreme
strong coupling limit, where the individual valence neutrons
do not alter the collective behavior (core behavior). On the other
hand the single particle contribution is sensitive to the number of
single particle states, but if more than the major shell is included,
this contribution to the
$BE(2)$'s converges (Fig.~\ref{fig:con}). We have also calculated
the $BE(2)$'s for different values of the effective charge.
The best fit is for $e_{{\rm eff}}$ close to one. However,
since the contribution
of the single particle term is very small, the  value of
$e_{{\rm eff}}$ is of little importance. The final results
of theoretical calculations for $^{155}$Gd and  $^{157}$Gd are
presented in figure \ref{fig:f1} and \ref{fig:f2} together
with experimental data and results of other theoretical models.

\begin{figure}
\centerline{\epsfysize=8cm \epsffile{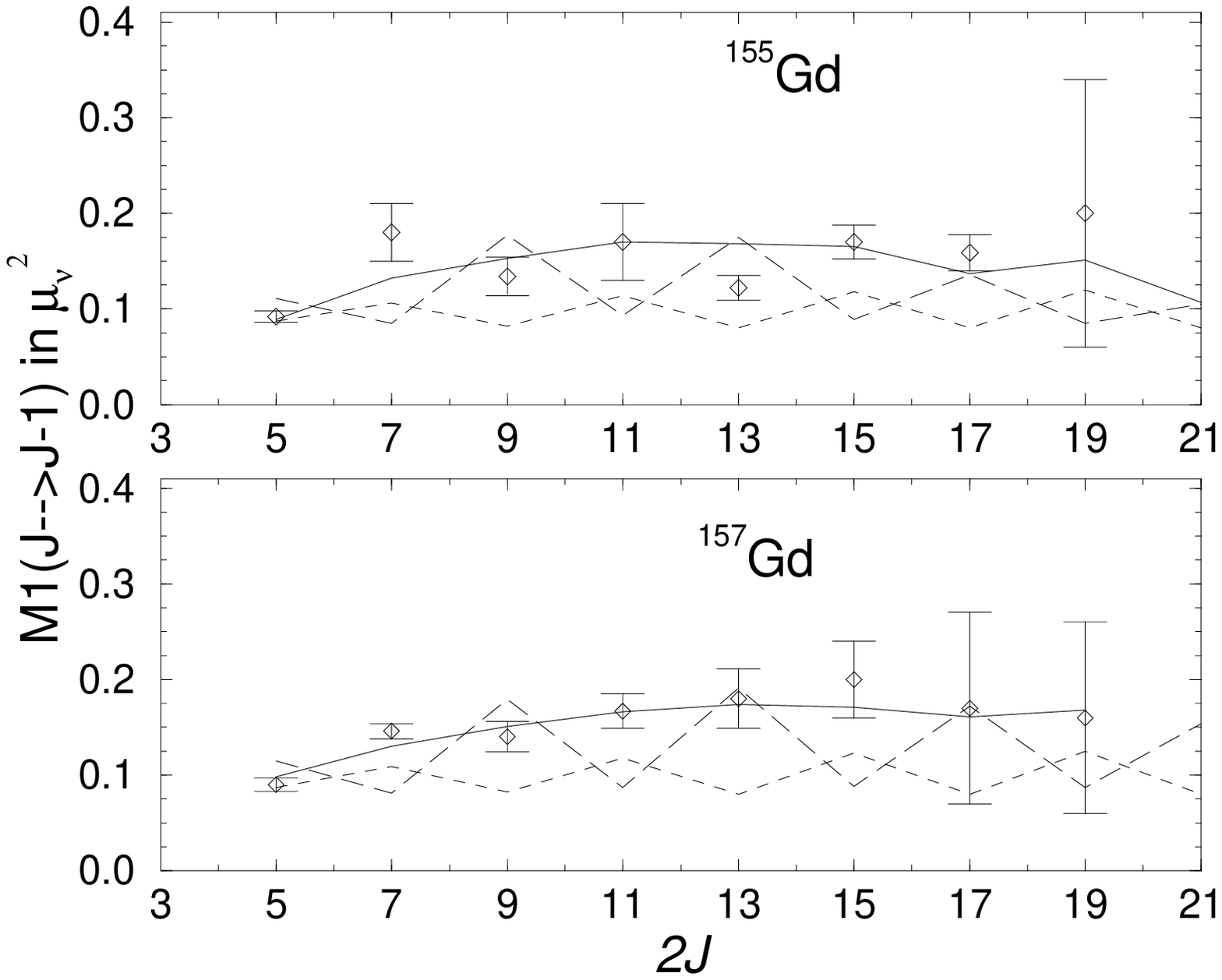}}
    \caption{ \label{fig:m1}
          The calculated values for the  M1 for $^{155,157}$Gd. The
parameters used are are listed in Table 1.
The points with error bars are the
experimental data \protect{\cite{e:1}}, the short-dashed line is the
single
particle contribution, the long-dashed line is the core contribution
and the
solid line is the total calculated $M(1)$.}
\end{figure}

We next calculate the magnetic transitions.
Here we have four parameters to fit. Because of the limitation on
available experimental data for
the magnetic moments of these states of $^{156,158}$Gd, we have to use
a phenomenological model for their values, \cite{fp:1}
\begin{equation}
	\left< I \| M \| I \right> = g_{1} I + g_{2} I^{2}
,
\end{equation}
where $g_{1}$ and $g_{2}$ are free parameters which we fit to the
experimental points.
In practice the experimental values permit a certain flexibility in the
choice of these parameters. The values found by fitting the
experiment in $^{156}$Gd and the values used are shown in
Table (\ref{t:m1}).
We further adjusted $g_{1}$, $g_{2}$ and $g_{S}$ slightly
 in order to find the best reproduction of the experiment
(see Tab. (\ref{t:m1})). Figure \ref{fig:m1} shows the calculations
for the
core contribution, the single particle contribution and the total $M1$.
As can be seen from the figure, the core and single particle
$M1$'s have roughly the same amplitude, and their amplitudes where
chosen
so as  to achieve the best total $M1$.
\begin{table}
\begin{center}
\begin{tabular}{ccccc}
quantity & theoretical & fit & $^{155}$Gd & $^{157}$Gd \\
\hline
$g_1$ & & $0.28 \pm 0.05$ & 0.27 & 0.3\\
$g_2$ & & $0.018 \pm 0.008$& 0.019& 0.02\\
$g_L$ & 0 & &    0 &  0\\
$g_S$ & -3.826 & & -3.586 & -3.576
\end{tabular}
\end{center}
\caption{\label{t:m1}
The values of the parameters used for the calculations of $M(1)$'s
(last two columns). The second column shows the theoretical values
for $g_L$ and $g_S$, whereas the third column shows the results of a
fit to $^{156}$Gd for the parameters  $g_{1}$ and $g_{2}$.}
\end{table}

\section{Review of previous Work}
We turn now to previous theoretical work on $^{155,157}$Gd, in order
 to summarize it  and compare it
with our own work and with experiment.
We can divide this work into three major groups according to
the formulation applied.

\begin{figure}
\centerline{\epsfysize=8cm \epsffile{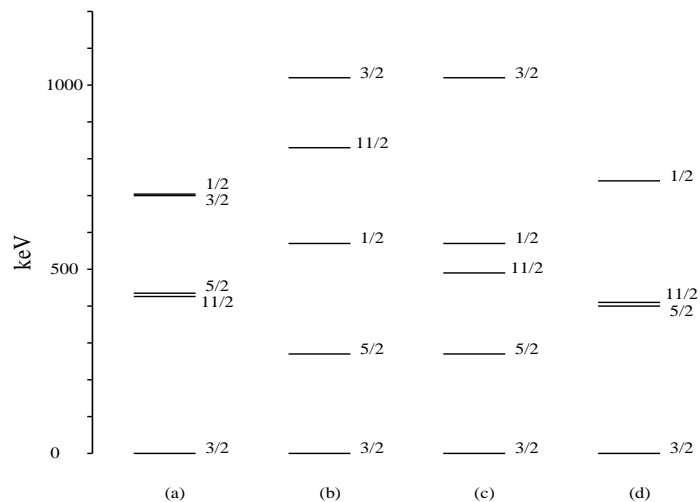}}
\caption{
   \label{fig:fig1} Negative parity band-head energies for $^{157}$Gd
from: (a) experiment, (b) reference \protect{\cite{a:1}}, (c)
reference \protect{\cite{a:2}},
(d) our work.}
\end{figure}

In the papers of the first  group, Refs. \cite{a:1,a:2,a:3,a:4}
only band-heads are calculated, using  the quasiparticle-phonon coupling
scheme. In this work, the underlying physics is that of
the standard core-particle coupling (CPC) model. Allowance is made for
phonon excitations of the core,
which can have both a
quadrupole and hexadecapole deformation.
The single particle states are found
using a Woods-Saxon
potential, and one uses a BCS approximation to include pairing
correlations.
In addition,
interactions between states with principle quantum number differing
by two are also included.
Only   band-heads
have been calculated with this model, and the results show little
improvement,
as regards experiment, over calculations based on the Nilsson model,
although fewer parameters are used in this approach \cite{a:3}.
In all these calculations the Coriolis interaction is neglected.
The argument is that the Coriolis interaction
has only a slight effect on the energies and structures of the lowest
parts of
the rotational bands \cite{a:1,a:2}. The results of the best two
calculations
of this type
are compared with experiment and with our calculations in Fig.~
\ref{fig:fig1}.

\begin{figure}
\centerline{\epsfysize=8cm \epsffile{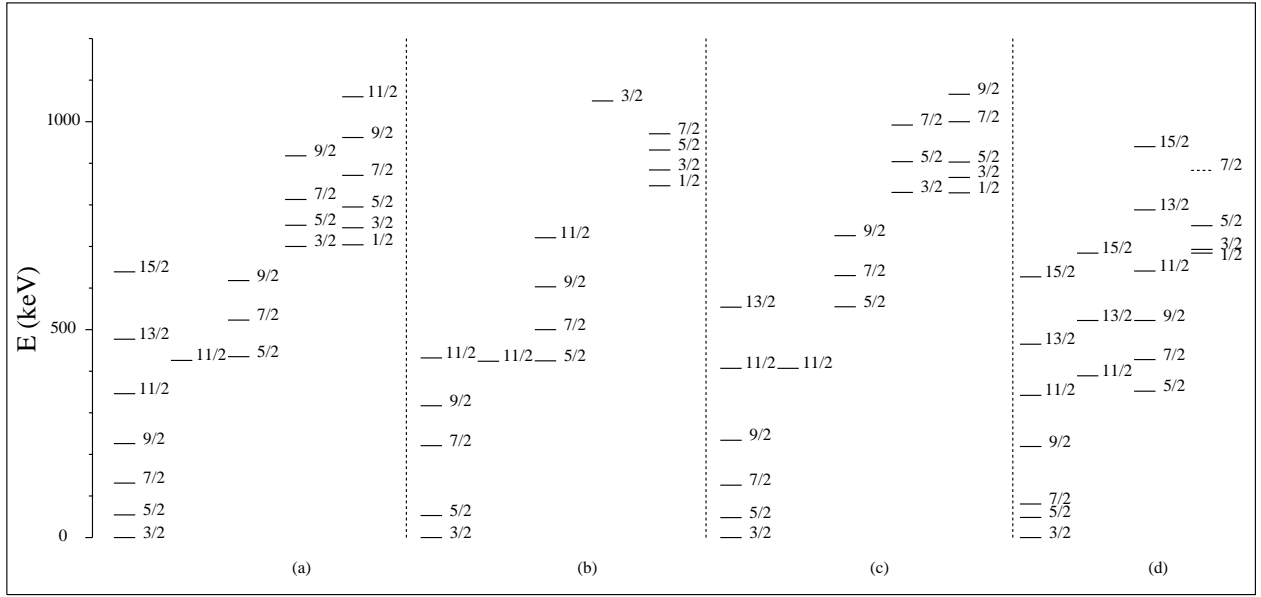}}
\caption{\label{fig:fig2} Negative parity states for $^{157}$Gd from:
(a) experiments, (b) reference \protect{\cite{b:1}}, (c) reference
\protect{\cite{b:2}}, (d) our work. }
\end{figure}

The second group, Refs.\cite{b:1,b:2,b:3,b:4} contains
complete calculations of the rotational spectrum in
the standard core-particle coupling model. Early works based on this
model utilize a rotating deformed core which is coupled to a
single odd nucleon. The mean field from the core is represented by
a Nilsson potential. The coupling between the core and the odd nucleon,
the Coriolis coupling, is treated as a perturbation (strong coupling
limit).
In Ref \cite{b:1,b:2} a deformed
Woods-Saxon potential is considered.
Again, pairing is taken into account by means of a BCS transformation.
Another extension of the original
model takes into account mixing between different major shells
\cite{b:1,b:2}.
In Fig.~\ref{fig:fig2} we show the experimental values compared to
three theoretical models. The first one (Fig.~\ref{fig:fig2}b) is from
\cite{b:1}. In this paper the model Hamiltonian includes a
quadrupole and hexadecapole deformed core. The pairing is treated
in BCS approximation. The deformation parameters $\beta_{20}$,
$\beta_{40}$ and the
moment of inertia ${\cal I}$ are taken as freely adjustable parameters.
The set of results shown in Fig.~\ref{fig:fig2}c is from
Ref. \cite{b:2}.
In this paper
the same Woods-Saxon potential and BCS approximation are used.
The model contains two additional free parameters:
the moment of inertia and the Coriolis attenuation factor $\kappa$,
which was found to be 0.63.
The last part of Fig.~\ref{fig:fig2} is based on our results, without
any Coriolis attenuation.


The third group includes models based on dynamical symmetry arguments;
the
pseudo SU(3) \cite{c:1} and the interacting boson-fermion approximation
(IBFA)
\cite{c:2}. The results are reasonable for both models but they will not
 be
reviewed because they describe only the lowest bands.

\section{Conclusions}
  In this paper we have applied to $^{155,157}$Gd the Core-Particle
model as originally formulated by Klein and co-workers and modified for
practical application by D\"{o}nau and Frauendorf.
When levels from several major shells were included, the
band-heads were predicted quite satisfactorily. The rotational structure
is predicted with high accuracy and only the higher bands are distorted.
We believe that the reason
for this distortion is the Coriolis interaction which is included
naturally
in the core Hamiltonian ($H_{c}$). The standard approach to
this difficulty is to attenuate the Coriolis interaction
\cite{b:1,b:2}. Since there is no justification for this ad-hoc
procedure,
the best approach is to include
other forms of interaction which will counteract this effect.
The suggested candidates are the quadrupole-pairing and
hexadecapole-hexadecapole interactions, which will
be investigated in the further development of this work.

The electric transitions $BE(2)$ were predicted and the results
are closer to the experimental values than any other theory for which
comparisons have been made.
For the magnetic transitions $M(1)$ we reproduce
the experiment but need 4  parameters to do so. These parameters
are not fully constrained by experiment.

Finally, it may be useful to summarize in general what has been
accomplished
and how the model may be extended. In our approach, the kinematical
Coriolis coupling between the particle and the core is automatically
included.
There is in fact no natural way of introducing an ad-hoc attenuation,
for which we have found little evidence. There
is some indication in higher excited bands, but not enough data is
 available
to substantiate this. In our calculations an excessive influence of
this coupling can be dealt with only by modifying the effective
interaction.
In addition non-adiabatic corrections
to the treatment of the core may be studied, either by use of
experimental matrix elements or by calculating higher order corrections
to the Bohr-Mottelson theory. Finally to study experimental results at
higher angular momentum, we must include excited bands of the core.
Because of the modest success of the calculations reported in this
paper, we feel encouraged to continue our investigations along
the lines suggested above.

\section{Acknowledgement}
This work was supported in part by U.S. Department of Energy
under grant number 40264-5-25351

\end{document}